\documentclass[twocolumn,superscriptaddress,aps,prl,longbibliography]{revtex4-2}
\usepackage{amssymb}
\usepackage{latexsym}
\usepackage{amsmath}
\usepackage[mathcal]{eucal}
\usepackage{tabularx}
\usepackage{bbold}
\usepackage{graphicx}
\usepackage{amsfonts}
\usepackage{comment}
\usepackage{bm}
\usepackage{epsfig}
\usepackage{array}
\usepackage[dvipsnames]{xcolor}
\usepackage{pifont}
\usepackage[normalem]{ulem}
\usepackage{standalone}

\begin{document}

\begin{abstract}
    We discuss a possibility of superconductivity in the $p$-wave magnets. These are recently discovered materials that have zero net magnetization by symmetry and finite non-relativistic spin splitting of electron bands, like in altermagnets. Similarly, the spin polarizations is collinear in the momentum space. Yet, as opposed to altermagnets, the magnetization is noncollinear in the real space, and the spin splitting obeys time-reversal symmetry in the momentum space. As a result, if such material harbors superconductivity (due to phonons, or any other mechanism), the only supported superconducting symmetry is Ising superconductivity, an exotic symmetry where any Cooper pair is a 50:50 mix of singlet and triplet. This unusual behavior is also in stark contrast to regular antiferromagnet, which can support Cooper pairs of any parity, and altermagnets, which can only support nonunitary triplet pairs. 
    The presence of large triplet component and enhanced resilience against pair breaking is inherent to the p-wave magnets and as such is unconventional as it does not materialize in conventional spin-orbit coupling induced Ising superconductors.
\end{abstract}
\title{Nonrelativistic-Ising superconductivity in p-wave magnets}

\author{M. Khodas}
\affiliation{Racah Institute of Physics, Hebrew University of Jerusalem, Jerusalem 91904, Israel}
\affiliation{Materials Science Division, Argonne National Laboratory, Lemont, Illinois 60439, USA}
\email{maxim.khodas@mail.huji.ac.il}

\author{Libor Šmejkal}
\affiliation{Max Planck Institute for the Physics of Complex Systems, 01187 Dresden, Germany}
\affiliation{Max Planck Institute for Chemical Physics of Solids, 01187 Dresden, Germany} 
\affiliation{Institute of Physics, Academy of Sciences of the Czech Republic, 162 00 Praha 6, Czech Republic}
\email{lsmejkal@pks.mpg.de}

\author{I. I. Mazin} 
\affiliation{Department of Physics and Astronomy, George Mason University, Fairfax, VA
22030, USA}
\affiliation{Quantum Science and Engineering Center, George Mason University, Fairfax, VA
22030, USA}
\date{\today}
\maketitle

\paragraph{Introduction ---}
$P$-wave magnets (pwM) are magnetic materials where the net magnetization is zero by symmetry, magnetic moments are collinear in momentum space, and the band splitting obeys time-reversal, $E_{\mathbf{k}\uparrow}-E_{\mathbf{k}\downarrow}=E_{\mathbf{-k}\downarrow}-E_{\mathbf{-k}\uparrow}$ \cite{Hellenes2024,Brekke2024,Chakraborty2025,Yu2025,Song2025,Yamada2025,Jungwirth2026}. 
A perfect, if trivial, example of the time-reversal symmetric spin split system is a nonmagnetic non-centrosymmetric 2D material with the Ising-type spin-orbit coupling (SOC) \cite{Lu2015,Ugeda2015,Saito2016,Xi2016,Costanzo2016,Sohn2018,Zhang2021,Sohier2025}, like NbSe$_2$ monolayer. ``Ising'' in this context means that electrons spins are 
collinear.
The spins are polarized out-of-plane due to a basal plane mirror symmetry. 
As a result, all 2D Fermi-surfaces, except at some special directions, are spin-split, and $E_{\mathbf{k}\uparrow}=E_{\mathbf{-k}\downarrow}$.
Such materials exhibit many unconventional properties, among the most intriguing of which is the so-called Ising superconductivity (IS), explained later in this Letter.

One limiting factor that prevents many unconventional phenomena predicted in Ising superconductors to manifest themselves in real materials is the fact that SOC is a relativistic effect, and, as such, very small in 3d metals, somewhat larger 4d, and moderate in 5d. Thus, all IS phenomena can be classified into two groups: those that require the SOC splitting $\xi_{\mathrm{SO}}$ to be stronger than the superconducting gap $\Delta$ (most importantly, protection from Pauli limiting), and those that require it to be comparable to the Fermi energy (or, in momentum space, that the spin splitting of the Fermi surface be comparable to the Fermi momentum $k_F$. The latter is not satisfied in known IS materials.

Here we show that this limitation can be circumvented in p$wM$s proposed few years ago \cite{Hellenes2024}. 
From the IS point of view, there are two fundamental differences between SOC-driven IS and non-relativistic, exchange-drive IS (NR-IS). 
First, while basically all practical implementations of SOC-IS are hexagonal, the $p$wMs can be orthorhombic. 
As a result, the minimal theoretical models for the former involve $f$-wave symmetry of the spin polarization and for the latter $p$-wave one.
Second, the exchange mechanism can generate  spin splittings on the order of eV \cite{Hellenes2024} replacing the order-of-magnitude weaker SOC. 
This leads not only to quantitative but also to qualitatively new phenomena, such as the non-relativistic Edelstein effect \cite{Chakraborty2025}, anomalous non-linear Hall effect \cite{Yamada2025}, net-triplet pairing symmetry, field induced non-unitary triplet pairing and reentrant superconductivity.
While SOC-IS beyond two spatial dimensions has been discussed only recently \cite{3D}, in NR-IS it comes naturally. 
And while in SOC-IS the spin are collinear only in the 2D plane, in NR-IS their collinearity is  protected in the full 3D BZ\cite{Hellenes2024}.

IS has been studied in NbSe$_2$ single layers and related compounds. 
The heart of the matter here is the bands crossing the Fermi level being composed of Nb $xy$, $x^2-y^2$ and $z^2$ orbitals, which can only form an orbital moment parallel to $z$ (out of plane). As a result, due to the absence of inversion symmetry, the Fermi surfaces split in two (apart from the nodal line $\Gamma$-M, where the bands are pure $z^2$ by symmetry), such that the spin and orbital moment in one valley is parallel, and in the other antiparallel to $\bm{z}$. 

The bands splitting is on the order of the SOC parameter $\xi_{\mathrm{SO}}$, such that $H\sim\Delta\ll\xi_{\mathrm{SO}}\ll E_F$, where $H$ is an external magnetic field.
Because of the first inequality, an in-plane field does not shift, in the lowest order, the Fermi surfaces, as opposed to the nonrelativistic case or to the relativistic case with $\bm{H}||\bm{z}$. 
Magnetization in this case is induced by canting the spins in the narrow belt between the two spin-split Fermi surfaces. 
It can be shown \cite{Wickramaratne2020} that the response is  the same, in the lowest order, as the Pauli response, even though the physical origin is different. 
Since this canting  mostly applies to electrons far (compared to $\Delta$) from the Fermi level, it is not affected by superconductivity, which ensures the absence of a thermodynamic Pauli limit. As we shall discuss later, it fully applies to NR-ISs as well.

Another ramification of the above is that each individual Cooper pair is either $|\bm{k},\uparrow; -\bm{k},\downarrow\rangle$ {\em or} $|\bm{k},\downarrow; -\bm{k},\uparrow\rangle$, as opposed to pure singlet superconductivity with each pair  described by 
$\Psi_s = |\bm{k},\uparrow; -\bm{k},\downarrow\rangle - |\bm{k},\downarrow; -\bm{k},\uparrow\rangle$, or by $\Psi_{t} = |\bm{k},\uparrow; -\bm{k},\downarrow\rangle +|\bm{k},\downarrow; -\bm{k},\uparrow\rangle$ in the pure triplet case. 
 In materials without inversion symmetry, like a NbSe$_2$ single layer, spin-singlet and spin-triplet pairs mix \cite{Gorkov2001,Yip2014,Samokhin2015}, so that the full superconducting order parameter can be written as $\Delta(\bm{k}) =\left(\psi_{\bm{k}}\sigma_0+\bm{d}_{\bm{k}}\cdot\boldsymbol{\sigma} \right)i\sigma_y$ with $\psi_{\bm{k}}$ and $\bm{d}_{\bm{k}}$ being singlet and triplet components, and $\sigma_0,\boldsymbol{\sigma}$ are the Pauli matrices parameterizing the spin space. 
 In SOC-ISs, as well as in NR-ISs,
 the amplitudes of spin singlet and spin triplet Cooper pairs, $\Psi_{s,t}$ are finite. 
 Unfortunately, due to the smallness of SOC in SOC-ISs, $\xi_{\mathrm{SO}}\ll E_F$, a  $|\bm{k},\uparrow; -\bm{k},\downarrow\rangle$ pair has a partner with nearly the same amplitude that is a $|\bm{k},\downarrow; -\bm{k},\uparrow\rangle$ pair, so the amplitude of the triplet state, for all practical purposes, is vanishingly small. {\em As discussed below, this is not the case for NR-ISs.}

For concreteness below we discuss a model of the $p$wM superconductor.
Still our conclusions hold for magnets of any odd symmetry including the recently discovered superconducting  iron based LaFeAs$_{0.6}$P$_{0.4}$ \cite{Stadel2022,Dsouza2026} which is an existing example of an $h$-wave magnet with magnetism driven by exchange interaction \cite{Yu2025}.

\paragraph{Effective model of $p$wM}

The model Bloch Hamiltonian describing the $p$wM given in the Ref.~\cite{Hellenes2024} (see also \cite{Mitscherling2026}) reads, 
\begin{align}\label{eff:model}
    H_{\bm{k}} & = 2 t (\cos \frac{k_x}{2} \tau_1 + \cos k_y) 
    \notag \\
    &+ 2 t_J (\sin \frac{k_x}{2}\tau_2 \hat{\bm{J}} \cdot \bm{\sigma}   + \cos k_y \tau_3 \hat{\bm{J}}' \cdot \bm{\sigma} )\, ,
\end{align}
where the in-plane magnetization unit vectors at magnetic sites are $\hat{\bm{J}} = \cos \phi \hat{x} + \sin \phi \hat{y}$ and $\hat{\bm{J}}' = \cos \phi' \hat{x} + \sin \phi' \hat{y}$.
The doubled unit cell contains two non-magnetic atoms belonging to  two distinct sublattices\cite{Hellenes2024}.
In the four dimensional space of Bloch states, $V^B_{\bm{k}}$, the $\sigma_i$ and $\tau_i$  are the Pauli matrices parameterizing the spin and the sublattice degrees of freedom, respectively. 

The coplanar structure is characterized by the antiunitary symmetry $\mathcal{T}[C_{2\perp}||E]$ that is the time reversal $\mathcal{T}$ followed by the two-fold spin rotation $C_{2\perp}$ in the plane defined by the magnetic moments.
Here $[o_s||o_l]$ denotes the operation $o_s$ acting on spins combined with the $o_l$ acting on the orbital degrees of freedom.
The system also has another antiunitary symmetry $\mathcal{T}\mathbf{t}$, where $\mathbf{t}$ is a translation by (half of the) basis vector of the (magnetic) non-magnetic lattice $ a \hat{x}/2$.
Thus, the model possesses a unitary symmetry, $[C_{2\perp}||\mathbf{t}]$.
Thanks to this symmetry the Bloch states at a generic $\bm{k}$ have zero  in-plane components of the spin polarization, $\langle \sigma_{1}\rangle = \langle \sigma_{2}\rangle =0$.
Stated differently, in terms of the model \eqref{eff:model},  $[C_{2\perp}||\bm{t}]$ is $S=\sigma_3 \tau_1$.
And vanishing of $\langle \sigma_{2,3} \rangle$ follows from $S^{-1}\sigma_{1,2} S = -\sigma_{1,2} $.
For $t=0$ the Hamiltonian \eqref{eff:model} anticommute with $\sigma_3$. 
Therefore, at $t=0$, and at $k_x = \pm \pi$ if there is no degeneracy, the expectation value of all  components of the spin operator vanish.

The $S = \tau_1 \sigma_3$ symmetry implies that the two sectors, $V_{\bm{k},S=\pm1}$ of Bloch states defined by $S=\pm 1$ decouple for all $\bm{k}$.
Namely, in the basis adjusted to the decomposition, $V^B_{\bm{k}} =V_{\bm{k},S=1} \oplus V_{\bm{k},S=-1}$ the Hamiltonian \eqref{eff:model} is block diagonal.
This is a generic feature of the NR models with the period doubling caused by the magnetic order \cite{Yu2025}. 
The $V_{\bm{k},S=+1} $ sector is spanned by the bonding spin up state, $\tau_1 = \sigma_3= +1$ and anti-bonding spin down state, $\tau_1 =\sigma_3 =-1$.
Similarly, $V_{\bm{k},S=-1} $ is spanned by the pair of states $\tau_1 = - \sigma_3 = \mp 1$.
Introducing the pseudo-spin Pauli matrices $\rho_{i}$, $i =0,1,2,3$ to parameterize the two sectors,  the two blocks of the Hamiltonian \eqref{eff:model} in $V_{\bm{k},S=\pm1}$ are 
\begin{align}\label{eff:model2by2}
    H_{\pm}(\bm{k}) = 2 t \cos k_y+ 2\bm{h}_{\pm}(\bm{k}) \cdot \bm{\rho}\, ,
\end{align}
\begin{align}\label{h_pm}
    [h_\pm (\bm{k}) ]_1 & =  t_J \left(\pm \sin \frac{k_x}{2}   \sin \phi +  \cos k_y \cos \phi' \right)
    \notag \\
    [h_\pm (\bm{k}) ]_2 & =  t_J \left(\mp \sin \frac{k_x}{2} \cos \phi + \cos k_y  \sin \phi'   \right)
    \notag \\
    [h_\pm (\bm{k}) ]_3 & =  \pm  t \cos \frac{k_x}{2} \, .
\end{align}
The spectrum and the spin polarization has been recently studied in details in Ref.~\cite{Mitscherling2026}.
For us it is important to realize that Cooper pairs are formed by partner states related by $\mathcal{T}\mathbf{t}$ symmetry.
However, as $(\mathcal{T}\mathbf{t})^{-1} S \mathcal{T}\mathbf{t} = -S$ the Cooper pairing combines electrons at different $S$-sectors.
Furthermore, the Cooper pair partners are the eigenstates of the pseudo-spin $1/2$.  
Although $\sigma_3 = \rho_3$,  $\sigma_{1,2} \neq \rho_{1,2}$ and in general the pseudo-spin states have to be projected on the interaction channels with well defined total spin.
The full analysis of the pairing is beyond the scope of this work.
Instead, we consider the  $t_J/t \ll 1$ regime to illustrate the generic features of pairing.

As is clear from Eq.~\eqref{h_pm} in this limit $\langle \rho_{1,2} \rangle/\langle \rho_3 \rangle \propto t_J/t \ll 1$ and the spin $\sigma_3 $ is approximately quantized.
At $t_J=0$ the eigenstates form two spin degenerate sectors $V_{\bm{k},\tau_1 = \pm 1}$,  split in energy by $\Delta E_k = 4 t \cos(k_x/2)$.
 Although $S$ has an advantage over the sublattice exchange parity $\tau_1$ of being exact symmetry, the Cooper pairs belong to the same parity and have opposite spins.
 Therefore, in the limit $t_J/t \ll 1$ it is more convenient to organize Bloch states by $\tau_1$.
 In the basis adjusted to the decomposition $V^B_{\bm{k}} =V_{\bm{k},\tau_1=1} \oplus V_{\bm{k},\tau_1=-1}$, Eq.~\eqref{eff:model} becomes
\begin{align}
    H_{\bm{k}} = \begin{pmatrix}
      2 t \cos k_y + \frac{\Delta E_{\bm{k}}}{2} & V_{\bm{k}} \\
        V^\dagger_{\bm{k}} & 2 t \cos k_y - \frac{\Delta E_{\bm{k}}}{2} 
    \end{pmatrix}\, ,
\end{align}
where $V_{\bm{k}} = 2 t_J (i  \sin\frac{k_x}{2} \hat{\bm{J}} \cdot \bm{\sigma} +  \cos k_y\hat{\bm{J}}' \cdot \bm{\sigma} )$.
With the leading order in $t_J/t$ correction, $V_{\bm{k}} V^\dagger_{\bm{k}}/\Delta E_{\bm{k}}$ \cite{Bjorken1964} included, the effective Hamiltonian in the $V_{\bm{k},\tau_1=1}$ subspace reads,
\begin{align}\label{eq:effective_gen}
    H^{\mathrm{eff}}_{\bm{k}} = a_{\bm{k}} + b_{\bm{k}} (\hat{\bm{J}}\times \hat{\bm{J}}') \cdot \bm{\sigma}\, ,
\end{align}
where $b_{\bm{k}} = - 2 t_J^2  \sin(k_x/2)\cos k_y / [ t \cos(k_x/2) ]$, and we ignore the corrections to the spin independent part, setting $a_{\bm{k}} = 2 t \cos k_y + 2t \cos (k_x/2)$.

As $\tau_1 =1$ is fixed the triple product in Eq.~\eqref{eq:effective_gen} is the only combination surviving the SU(2) spin transformations applied to both localized and itinerant moments allowed at zero SOC. 
The spatial inversion is broken by magnetism \cite{Hellenes2024,Lee2025}.
At $\hat{\bm{J}} \perp \hat{\bm{J}}'$ the spatial inversion becomes a symmetry when combined with the spin transformation restoring the original magnetic arrangement. 
Unlike the more familiar inversion in SOC-ISs this combined operation flips $\langle \bm{\sigma} \rangle$.
Therefore, in NR-ISs this feature does not affect superconductivity in the same way as in SOC-ISs and we set $\hat{\bm{J}} \times \hat{\bm{J}}' = \hat{z}$ for simplicity.

\begin{figure}[t!]
    \includegraphics[width=\columnwidth]{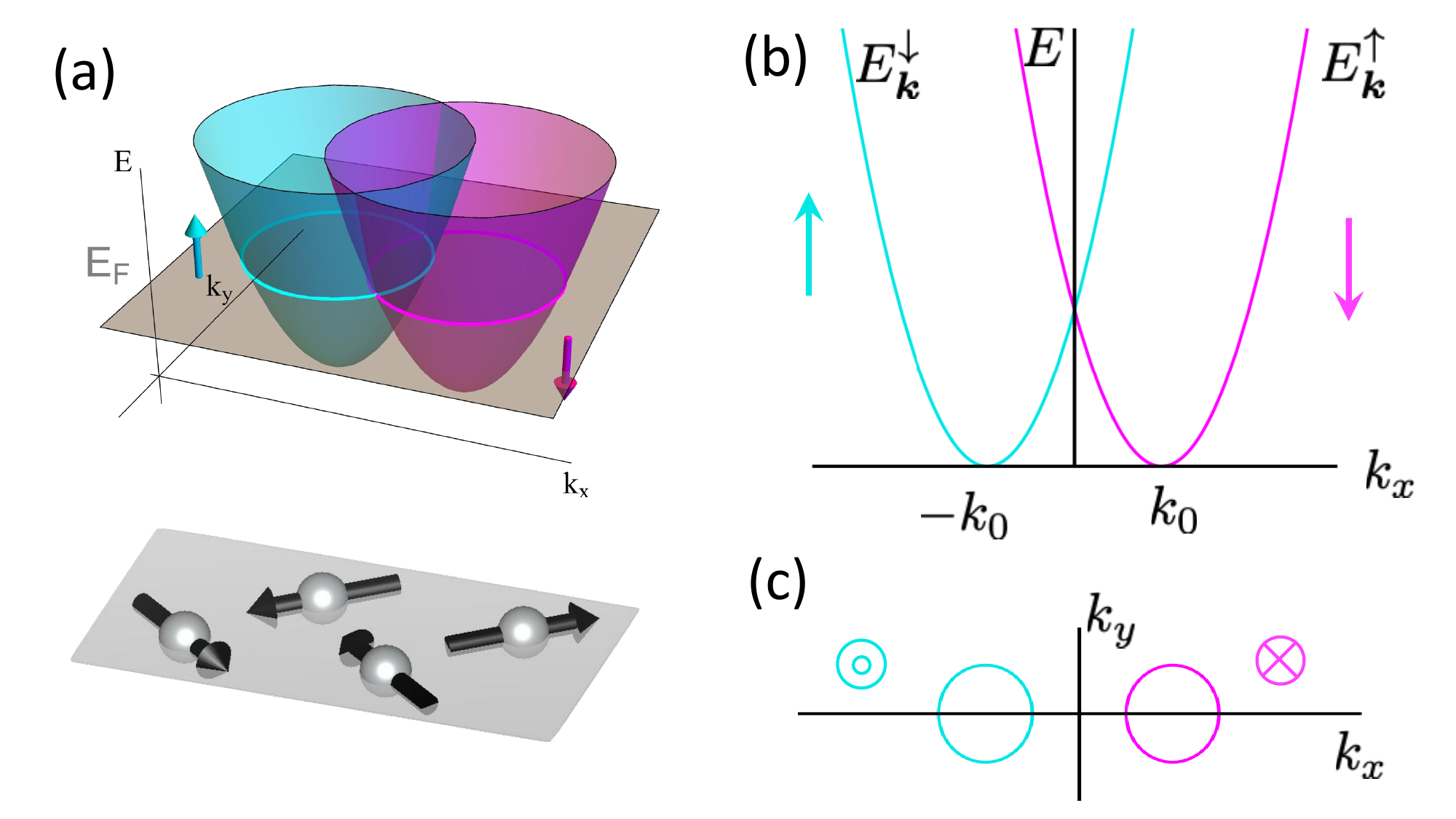}
    \caption{(a) The p-wave spin-split bands with the momentum space spins (top panel) oriented perpendicular to the direct space (bottom panel) noncollinear magnetic order. (b)
    The dispersion relations of the two spin split bands $E_{\bm{k}}^{\uparrow,\downarrow}$, Eq.~\eqref{eq:EkUD} at $k_y=0$ as a function of $k_x$. (c)
    At a fixed Fermi energy, $E_F$ the two oppositely spin polarized Fermi surfaces are disjoint at sufficiently large separation, $2k_0$.}
    \label{fig:dispersion}
\end{figure}

In the long wavelength limit the effective Hamiltonian,
\begin{align}\label{eq:model}
    H_{\bm{k}}^{\mathrm{eff}} = 
    \frac{k_x^2}{2m} - \frac{k_0 k_x}{m}\sigma_z
   + \frac{2}{m}k_y^2 + H \sigma_x \, ,
\end{align}
where $k_0 = -2 t_{J}^2/  t^2$, and $1/m = - t/2$. 
The exchange field $H\hat{x}$ we introduced as an addition to Eq.~\eqref{eff:model} breaks  $\mathcal{T}\mathbf{t}$ and $[C_{2\perp}||\mathbf{t}]$, yet crucially preserves their product, $\mathcal{T}[C_{2\perp}||E]$.
Without loss of generality, we apply Eq.~\eqref{eq:model} with the penultimate term replaced by $k_y^2/ 2m$, rendering the problem isotropic.
The most important feature of Eq.~\eqref{eq:model} is the momentum-dependent non-relativistic spin splitting,
$\xi_{\mathrm{nr}}(\bm{k}) =  2 k_0 k_x /m$.
At $\bm{H}=0$ the spin up and down subbands,
\begin{align}\label{eq:EkUD}
    E_{\bm{k}}^{\uparrow,\downarrow}=[ \left(k_x \mp k_0 \right)^2 + k_y^2]/2m\, ,
\end{align}
where we have added a constant to complete the expression for $E_{\bm{k}}^{\uparrow,\downarrow}$ to the square, see Fig.~\ref{fig:dispersion}.

At the Fermi energy, $E_F$, the spin up (down) Fermi surfaces form spheres shifted by $\pm k_0 \hat{x}$. 
For $k_F = \sqrt{2 m E_F} > k_0$ the two Fermi spheres intersect (case 1), otherwise, they are disjoint (case 2). 
For phenomena such as the Edelstein effect \cite{Edelstein1990,Chakraborty2025,Sun2025} and in-plane critical field anisotropy \cite{Wickramaratne2021,Haim2022} nodes play an important role, and this makes the case 1 more appropriate model.

The Cooper pairs with momenta at the crossings of the two Fermi surfaces (see Fig.~\ref{fig:dispersion}a) form pure singlets, as the triplet amplitude is odd in $k_x$.
Therefore, these specific Cooper pairs are affected by the magnetic field.
The expected topological transition to the nodal state \cite{He2018,Shaffer2020} has no bearing on the thermodynamic properties of ISs \cite{Kuzmanovic2022}.
It is therefore, justified and useful to consider the extreme case 2, $E_F \ll t_J^2/t$ of well-separated Fermi surfaces in the discussion of the properties of the order parameter defined globally across the Brillouin Zone.
In case 2, $\xi_{\mathrm{nr}}(\bm{k}) \approx \xi_{\mathrm{nr}}  \mathrm{sgn}(k_x)$, 
$\xi_{\mathrm{nr}} = 2 (k_0^2 /m)$.
For clarity, we consider a weakly bound Cooper pairs, $\Delta \ll E_F$.
The results hold qualitatively regardless of the microscopic details such as the symmetry of the interaction channel, the size of the Cooper pair and the separation $k_0$ of the two Fermi surfaces.

\paragraph{The $s+p$ pairing in $p$wM}

We turn to the structure of the Cooper pairs unique to $p$wM in the extreme case 2, where the spin splitting is large for all momenta, 
while for intersecting Fermi surfaces the same is true for the overwhelming majority of the states at the Fermi level, setting for now $\bm{H}=0$.
An earlier paper \cite{Sukhachov2025} has considered a specific model for $p$wM, considering an attractive interaction in the singlet channel. 
Even if an attractive interaction is assumed to exist only in the singlet channel, the $p$-wave band structure inevitably leads to emergent triplet correlations manifest in the anomalous Green function.
For this reason, the singlet and triplet channels cannot be treated independently.

The triplet component of the anomalous average inherits the $p$-wave symmetry of the underlying band structure of the $p$wM.
Hence, one can also refer to this combined order as $s+p$.
To put it in correspondence with SOC-ISs we view the effective model \eqref{eff:model} as having the arising from orthorhombic $D_{2h}$ symmetry at $k_0=0$ broken down to $C_{2v}$ at finite $k_0$ (see Tab.~\ref{tab:D2h}).
This symmetry reduction eliminates the inversion, and preserves the basal mirror as is required for the Ising SOC.

As the spin dependent term $\propto k_0$ plays the role of the SOC we follow a well-known procedure to identify the triplet order parameter coexisting with the singlet one \cite{Yip1993,Yip2014}. 
The $A_{g}$ and $B_{2u}$ irreps of $D_{2h}$ merge in $C_{2v}$, see Tab.~\ref{tab:D2h}.
As a result, the singlet component coexists with the $\bm{d}_{\bm{k}} = \eta \hat{z} f_{\bm{k}}^t$, where $f_{\bm{k}}^t = k_x$ to the leading order in $\bm{k}$.
More generally, $f_{\bm{k}}^t$ is momentum odd and satisfies $f^t_{-k_x,k_y} = - f^t_{k_x,k_y}$.
The $p$-wave symmetry of the triplet amplitude $f^t_{\bm{k}}$ reflects the symmetry of the $p$-wave symmetric spin texture  of $p$wM, see Fig.~\ref{fig:dispersion}.
Similarly, in hexagonal SOC-IS the reduction of the $D_{6h}$ symmetry down to $D_{3h}$ causes the admixture of the opposite spin triplets of a similar form, $\bm{d}_{\bm{k}}= \hat{z} \eta f^t_{\bm{k}}$, 
where $f^t_{\bm{k}}$ has a three- rather than two-fold rotational symmetry, see Tab.~\ref{tab:D6h}.

A crucial difference between the SOC-ISs and NR-ISs is the relative admixture of the $\bm{d}\parallel \hat{z}$ triplets.
Although in both cases the singlet and triplet order parameters are allowed to mix by symmetry,
at $(\xi_{\mathrm{SO}}/E_F)^2 \ll 1$, they effectively decouple \cite{Gorkov2001,Frigeri2004,Frigeri2006} in stark contrast to Eq.~\eqref{eq:self1}. 
One therefore can speak of a (nearly) pure singlet superconductivity in SOC-ISs.
 In contrast, in NR-ISs the  singlet and $\bm{d}\parallel \hat{z}$ triplet always occur together and one cannot be considered without the other.

\begin{table}
    \centering
    \begin{tabular}{|c|c||c|c|c|c|}
    \hline
            $\bm{H}$   &   $C_{2v}$& $D_{2h}(e)$ & $\psi_{\bm{k}}$ & $D_{2h}(o)$ & $\bm{d}_{\bm{k}}$ \\
    \hline
                &   $A_{1}$ & $A_{g}$ & $1$ & $B_{2u}$ & $\hat{x}k_z,\hat{y}k_x k_y k_z$, $\left[\hat{z}k_x\right]$   \\
        \hline
        $H_y$   &   $A_{2}$ & $B_{2g}$ & $k_x k_z$ & $A_u$ & $\left\{\hat{x} k_x\right\}$, $ \hat{y} k_y,\hat{z} k_z $  \\
        \hline
        $H_x$   &   $B_{1}$ & $B_{3g}$ & $k_y k_z$ & $B_{1u}$ & $\hat{x} k_y$, $\left\{\hat{y}k_x \right\}$, $\hat{z}k_x k_yk_z$  \\
        \hline
        $H_z$        &   $B_{2}$ & $B_{1g}$ & $k_x k_y$ & $B_{3u}$ & $\hat{x}k_xk_yk_z,\hat{y}k_z,\hat{z}k_y$ \\
        \hline
    \end{tabular}
    \caption{
    The $D_{2h}$ symmetry of the model Eq.~\eqref{eff:model} at $k_0 =0$  reduced to $C_{2v}$ at $k_0 \neq 0$. 
    The third and fifth columns list the even (e) and odd (o) irreps of $D_{2h}$ along with the scalar and vector functions to the leading order in $\bm{k}$, respectively. 
    The even and odd irreps of $D_{2h}$ merge pairwise into one irrep of $C_{2v}$ listed in the same line of the second column.
    In the 2D system, $k_z=0$ the $\bm{d}_{\bm{k}} = \hat{z}k_x$ in square brackets coexists with  singlet Cooper pairs in $C_{2v}$. 
    The $\bm{d}_{\bm{k}}$-vectors in the braces are induced by the field components shown in the same line.
    They are distinguished by the same $\bm{k}$-dependence ($\propto k_x$) as the triplet in the square brackets in the second line.}
    \label{tab:D2h}
\end{table}
\begin{table}
    \centering
    \begin{tabular}{|c|c||c|c|c|c|}
    \hline
            $\bm{H}$ &  $D_{3h}$ & $D_{6h}(e)$ & $\psi_{\bm{k}}$ & $D_{6h}(o)$ & $\bm{d}_{\bm{k}}$  \\
    \hline
                     &       $A'_{1}$ & $A_{1g}$ & $1$ & $B_{1u}$ & $\left[\mathrm{Im}k_+^3 \hat{z}\right]$, $\mathrm{Im}k_+^2k_z\hat{r}_+$     \\
        \hline
            $H_z$       &     $A'_{2}$ & $A_{2g}$ & $\mathrm{Im} k_+^6$ & $B_{2u}$ & $\mathrm{Re}k_+^3 \hat{z}$  \\
        \hline
                &   $A''_{1}$ & $B_{1g}$ & $\mathrm{Im} k_+^3 k_z$ & $A_{1u}$ & $\hat{z} k_z$, $k_x \hat{x}+k_y \hat{y}$  \\
        \hline
                 &  $A''_{2}$ & $B_{2g}$ & $\mathrm{Re} k_+^3 k_z$ & $A_{2u}$ & $\mathrm{Im}k_-\hat{r}_+$  \\
        \hline
               $H_{\pm}$   & $E''$     & $E_{1g}$ & $\pm k_z k_{\pm}$ & $E_{2u}$ &  
               $\{k_{\mp}^3\hat{r}_\pm\}$, $\{k_{\pm}^3\hat{r}_\pm\}$  \\
                &   & & &  & $k_{\mp} \hat{r}_{\mp}$, $k_z k_\mp^4 \hat{z}$  \\
        \hline
                 &  $E'$     & $E_{2g}$ & $k_{\pm}^2$ & $E_{1u}$ & $\pm k_{\mp} \hat{z}$, $\pm k_{z} \hat{r}_{\mp}$  \\
        \hline        
    \end{tabular}
    \caption{
    The same as Tab.~\ref{tab:D2h} for the hexagonal SOC-ISs, with the $D_{6h}$ symmetry   reduced to $D_{3h}$ lacking an inversion center still having the basal mirror plane \cite{Yip1993,Yip2014}. Here $k_\pm = k_x \pm i k_y$, $\hat{r}_{\pm} = \hat{x} \pm i \hat{y}$, $H_\pm = H_x \pm i H_y$.
    }
    \label{tab:D6h}
\end{table}

The above symmetry considerations allows us to formulate the minimal pairing Hamiltonian featuring all  pairing channels that cannot be disentangled in NR-IS,
\begin{align}
    H_p = g_s P_s^{\dagger} P_s + g_t  P_t^{\dagger} P_t ,
\end{align}
where  $P_{s} = \sum_{\bm{k}} f^{s}_{\bm{k}} c_{-\bm{k}s} [ i\sigma_{y(x)}]_{ss'} c_{\bm{k}s'}$ and 
$P_{t} = \sum_{\bm{k}} f^{t}_{\bm{k}} c_{-\bm{k}s} [ \sigma_{x}]_{ss'} c_{\bm{k}s'}$, and we set $f^{s}_{\bm{k}} = 1$.

The self-consistency conditions are 
\begin{align}\label{eq:self}
    \frac{\psi}{g_s}  = \sum_{\bm{k}} \mathrm{Tr} [i \sigma_y \hat{F}_{\bm{k}}(0) ] \, ,
     \frac{\eta}{g_t} = \sum_{\bm{k}} f^t_{\bm{k}}\mathrm{Tr} [\sigma_x \hat{F}_{\bm{k}}(0) ],
\end{align}
where the imaginary time ($\tau$) anomalous Green function $F_{\bm{k},ss'}(\tau) = - \langle T_{\tau}c_{\bm{k}s}(\tau)c_{-\bm{k}s'}\rangle$, where $c_{\bm{k}s}$ annihilates an electron, with momentum $\mathbf{k}$ and spin $s =  \uparrow,\downarrow$.
It is the off-diagonal block of the mean-field Green function, 
$\mathcal{G} (\bm{k},i \omega_n) = (i \omega_n - \mathcal{H})^{-1}$, and the Bogoliubov-de Gennes (BdG) Hamiltonian is,
\begin{align}\label{eq:BdG}
   \mathcal{H} =
    \begin{bmatrix}
         \xi^{\uparrow}_{\bm{k}} & 0 & 0 &- \psi - \eta f^t_{\bm{k}} \\
       0 & \xi^{\downarrow}_{\bm{k}} & \psi - \eta f^t_{\bm{k}} & 0 \\
        0 & \psi - \eta f^t_{\bm{k}}& -\xi^{\uparrow}_{-\bm{k}} & 0 \\
        -\psi - \eta f^t_{\bm{k}} & 0 & 0 &  -\xi^{\downarrow}_{-\bm{k}}
    \end{bmatrix},
\end{align}
where $\xi^{\uparrow,\downarrow}_k = E^{\uparrow,\downarrow}_k - E_F$, and and $\psi$ ($\eta f^t_{\bm{k}}$) are the singlet (triplet) components of the order parameter.

The BdG Hamiltonian \eqref{eq:BdG} makes it clear that the only Cooper pairs surviving the large non-relativistic spin splitting are $|\bm{k},\uparrow;-\bm{k},\downarrow \rangle$ for $k_x > 0$ and $|\bm{k},\downarrow;-\bm{k},\uparrow \rangle$ for $k_x < 0$.
The anomalous Green function is
\begin{align}\label{eq:F}
    F_{\bm{k},\uparrow\downarrow}(i \omega_n)& = \theta^>_{\bm{k}}\frac{\psi + \eta f_{\bm{k}}^t}{\omega_n^2 + (\xi_{\bm{k}}^\uparrow)^2+ (\psi + \eta f_{\bm{k}}^t)^2 }
    \notag \\
     F_{\bm{k},\downarrow\uparrow}(i \omega_n)& = \theta^<_{\bm{k}}\frac{-\psi + \eta f_{\bm{k}}^t}{\omega_n^2 + (\xi_{\bm{k}}^\downarrow)^2+ (\psi - \eta f_{\bm{k}}^t)^2 }\, ,
\end{align}
where $\theta^{>(<)}_{\bm{k}} = 1$ for $k_x> 0$ ($k_x < 0$) and zero otherwise.
At a given $\bm{k}$ either $F_{\bm{k},\uparrow\downarrow}$ or $F_{\bm{k},\downarrow\uparrow}$ is non-zero.
Reflecting again, the equal mixture of the singlet and triplet components of the Cooper pair wave function, $\hat{F}_{\bm{k}}$. 
Note that as $E_{\bm{k}}^{\uparrow} = E_{-\bm{k}}^{\downarrow}$, the Cooper pair wavefunction, \eqref{eq:F} satisfy, $F_{\bm{k},\uparrow\downarrow} = - F_{-\bm{k},\downarrow\uparrow}$ as prescribed by the Pauli principle.

Setting $ f^t_{\bm{k}} = \mathrm{sgn}(k_x)$, and combining Eqs.~\eqref{eq:F} and \eqref{eq:self} we obtain at a temperature $T$,
\begin{align}\label{eq:self1}
   \frac{\psi}{g_s} = \frac{\eta}{g_t}  =  \pi N T \sum_{i \omega_n} \frac{\psi +\eta}{\sqrt{\omega_n^2 +(\psi +\eta)^2 }},
\end{align}
where $N$ is the density of states.
Equation \eqref{eq:self1} demonstrates the strong coupling between the two order parameters.

Keeping, close to the transition,  the terms up to the second order in $\psi$ and $\eta$, the free energy is
\begin{align}\label{F2_4}
    F = &  \frac{\psi^2}{g_s} +\frac{\eta^2}{g_t}-  N \log\frac{c \omega_D}{T} \left[ (\psi^2 + |\eta|^2 + 2 \psi \eta ) \right],
\end{align}
where $\omega_D$ is the pairing interaction energy scale  and $c$ is a constant of the order one.
Similar to Eq.~\eqref{eq:self1}, Eq.~\eqref{F2_4} implies $\psi/g_s = \eta/g_t$ for any $g_t$ including $g_t =0$ i.e., $\eta=0$ in the latter case.
The constrain on the Cooper pairs to be half singlet and half triplet enforces the critical temperature of $T_c \propto \omega_c \exp[ (g_s + g_t)^{-1}N ^{-1} ]$ even in the presence of the repulsion in the triplet channel \cite{Dalal2023,Fanfarillo2025}.
As a consequence, as $g_t$ approaches $-g_s$ from below, the $T_c$ goes to zero.

\paragraph{Superconductivity beyond Pauli limit ---}
The $p$wM enjoys similar or higher level of protection against the pair breaking \cite{Bulaevskii976,Ilic2017}.
Similar to IS, in the $p$wM the spin susceptibility is not affected by the superconductivity. 
In fact, except for a tiny fraction, all the electrons contribute to the magnetization by adjusting theirs spins exactly as in the normal state.

The above argument generally applies in the $T \ll T_c$ limit.
In Ref. \cite{Ilic2017} the opposite limit, 
$T_c - T \ll T_c$ was considered, and, under an implicit assumption that  $\xi_{\mathrm{SO}}\ll \omega_D$, an expression for the  critical field 
$H_c = \xi_{\mathrm{SO}} [\log( \xi_{\mathrm{SO}}/2 \Delta)]^{-1/2} ( 1 - T/T_c)^{1/2}$ was obtained. In reality, even in NbSe$_2$, $\xi_{\mathrm{SO}}\gtrsim$ 100 meV, while $\omega_D\sim 30$ meV. Obviously, NR-ISs are also in this regime, where similar arguments lead to another logarithmically-accurate estimate,
$H_c= \xi_{\mathrm{nr}} [\log( \omega_{D}/2 \Delta)]^{-1/2} ( 1 - T/T_c)^{1/2}$.
As $\xi_{\mathrm{nr}} \gg \xi_{\mathrm{SO}}$ the parallel critical field of the NR-SOC is larger than the critical field of IS by a factor $\xi_{\mathrm{nr}} / \xi_{\mathrm{SO}} \gg 1$.

\paragraph{Field-induced $s+ p + i p'$ superconductivity ---}
%
\begin{figure}[ht]
    \includegraphics[width=0.95\columnwidth]{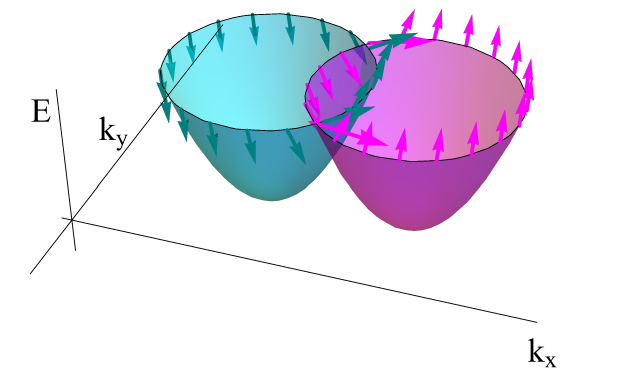}
    \caption{
    The dispersion relations of the two spin split bands are not modified by the exchange field, $\bm{H} = H \hat{x}$ at $k_y=0$ in the lowest order in $H$, but spins cant into the field direction.
    The states at the two branches with opposite momenta are related by the $\mathcal{T}[C_{2\perp}||E]$ symmetry.
    At large enough $E_F$ the two branches remain disjoint at finite exchange field as in Fig.~\ref{fig:dispersion}.
    }
    \label{fig:dispersion1}
\end{figure}
In NR-ISs, the in-plane exchange field respects the $\mathcal{T}[C_{2\perp}||E]$ symmetry connecting the electrons forming a Cooper pair.
The exchange field transforms the pair wave function by canting the spins \cite{Mockli2019,Mockli2022,Ilic2023}.
Based on the considerations on SOC-IS \cite{Mockli2019,Fischer2018,Mockli2022,Kuzmanovic2022,Ilic2023} we expect a transformation of the $s+p$ in $p$wM.

The mixed $s+p$-wave  order parameter of a $p$wM responds to the exchange field in a way that is qualitatively distinct from the $s$-wave order parameter of SOC-ISs.
To understand this difference it is instructive to follow the evolution of a $|\bm{k}\uparrow;-\bm{k}\downarrow \rangle$ Cooper pair wave function with the in-plane field, see Fig.~\ref{fig:dispersion1}.
Let $\hat{U}_{\bm{n}\varphi}$ denote the spin rotation by an angle $\varphi$ around $\bm{n}$.
The  $\bm{H} = H \hat{x}$ field turns the Cooper pair wave function $\Psi$ into 
\begin{align}\label{eq:transform}
\Psi'& =    |\bm{k}, U_{\hat{y}\theta}\uparrow;-\bm{k}, U_{\hat{y},-\theta} \downarrow \rangle  
   \notag \\
 = &     |\bm{k} , \cos \theta/2 \uparrow + \sin \theta/2 \downarrow; -\bm{k}, \sin \theta/2 \uparrow + \cos \theta/2 \downarrow \rangle
    \notag \\
    = & 2 \sin \theta ( |\bm{k} ,\uparrow; -\bm{k} ,\uparrow \rangle + |\bm{k} ,\downarrow; -\bm{k} ,\downarrow \rangle) 
   \notag \\
     +  2  &( 1\!-\! \cos \theta) |\bm{k} ,\downarrow; -\bm{k} , \uparrow \rangle \!+\! 2 (1 \!+\! \cos \theta ) |\bm{k} ,\uparrow; -\bm{k} ,\downarrow \rangle,
\end{align}
where the angle $\theta$ satisfies
$\cos \theta =\xi_{\mathrm{nr}} \mathrm{sgn}(k_x) /(\xi_{\mathrm{nr}}^2 +H^2)^{1/2}$, $\sin \theta >0$.
Equivalently,  
\begin{align}\label{F2_6}
    \Psi' = i \sigma_y \cos \theta + \sigma_0 \sin \theta + \sigma_x\, .
\end{align}
To highlight the significance of Eq.~\eqref{F2_6} we, again, turn to the Landau free energy. 
In addition to the spin triplet order parameter $\bm{d}^R_{\bm{k}} =  \eta \hat{z}  k_x$ present at $H=0$ the exchange field makes it necessary to introduce the imaginary order parameter $\bm{d}^I_{\bm{k}} = - i \zeta \hat{y}  k_x$ that is a different $p$-wave ($p'$-wave) describing the parallel spin Cooper pairs.
The Landau functional accommodating both types of triplets extends Eq.~\eqref{F2_4} to  
\begin{align}\label{F2_5}
    F^{(2)}[\psi,\eta,\zeta] = & 
    |\psi|^2/g_s  + \eta^2/g_t + \zeta^2/g'_t \notag \\
    + 2N & \log\frac{c \omega_D}{T} (\zeta + \psi \cos \theta + \eta \sin \theta)^2, 
\end{align}
resulting in coupled linearized self-consistency equations, 
\begin{align}\label{F2_7}
    \psi/g_s & = -2 \log( c \omega_D/T)    (\zeta + \psi \cos \theta + \eta \sin \theta) \cos \theta 
    \notag \\
    \eta/g_t & = -2 \log( c \omega_D/T)  (\zeta + \psi \cos \theta + \eta \sin \theta) \sin \theta 
    \notag \\
    \zeta/g'_t & = -2 \log( c \omega_D/T)  (\zeta + \psi \cos \theta + \eta \sin \theta) \, . 
\end{align}
in agreement with Eq.~\eqref{F2_6}.
While the $s+if$ in hexagonal SOC-ISs is unitary, $s+p+ip'$ in orthorhombic NR-ISs such as $p$wM is non-unitary.

It is worth noting that all SOC-ISs discussed in the literature are hexagonal, while nearly all $p$wM discussed so far  are orthorhombic. 
This is reflected in the above result, in the sense that the triplet component in the former is $f$-wave and in the latter $p$-wave. In addition, the large amplitude of the spin-splitting makes it necessary in the NR-IS case to add a $p'$ component rendering the superconductivity non-unitary.

\paragraph{Re-entrant field-induced superconductivity ---}
In ISs the pair breaking effect of magnetic impurities depends on the size $R_{\mathrm{imp}}$ of the impurities and on orientation of their magnetic moment, $\bm{m}$ \cite{Mockli2020}.
For $\bm{m}\parallel \hat{z}$, the impurities suppress $T_c$ through spin-conserving scattering.
For $\bm{m}\perp \hat{z}$, the scattering flips the electron spin.
Therefore, the scattering is suppressed if $R_{\mathrm{imp}} \gg v_F/\xi_{\mathrm{nr}(\mathrm{SO})}$ in NR-IS and OC-IS, respectively.
In a standard SOC-IS, $\xi_{\mathrm{SOC}}$ is quite small, and therefore the condition $R_{\mathrm{imp}} \gg v_F/\xi_{\mathrm{SO}}$ can rarely be satisfied.
To the contrary, the condition $R_{\mathrm{imp}} \gg v_F/\xi_{\mathrm{nr}}$ is very realistic.
As a result, in the NR-IS an applied field that rotates $\bm{m}$ from $\bm{m}\parallel \hat{z}$ to $\bm{m}\perp \hat{z}$ restores the clean-limit value of $T_c$. 
The superconductivity induces magnetization reorientation if the condensation energy is larger than magnetic anisotropy energy. 
Otherwise, the polarization direction can switch superconductivity on and off \cite{Fang2024}.

\paragraph{Conclusion ---}
In this Letter we have examined how the effects characteristic of ISs with the relativistic SOC manifest themselves in $p$wMs with the much larger non-relativistic spin splitting.
Some of the previous results apply equally to both systems.
These include the lack of Pauli limiting, field induced topological transition to the nodal state \cite{He2018,Chen2019}, field induced Bogoliubov Fermi surfaces \cite{Shaffer2020}, mirage gaps \cite{Tang2021} and critical in-plane anisotropy \cite{Hamill2021,Cho2022} all predicted for SOC-ISs.

In all ISc relativistic and non-relativistic alike the Cooper pair wave function for a given electron momentum is 50:50 mixture of singlet and triplet components.
However, while in the BCS models of SOC-ISs this is largely inconsequential in terms of the resulting order parameter, in NR-ISs it translates into the strongly coupled and distinct singlet and triplet components of the order parameter.
In particular, in orthorhombic $p$wM the transition temperature is set by the net interaction in the singlet and $p$-wave triplet channels. 

We note that the unconventional parity mixed superconductivity has been reported in recent studies of the SOC-ISs 
with multi-valley Fermi surfaces and realistic electron-phonon and Coulomb interactions \cite{Horhold2023,Sohier2025,Roy2025}.
We expect the unconventional character of these pairing states to be substantially more pronounced in NR-ISs.
The non-unitary and nodal pairing can be accessed via the quasi-particle interference \cite{Hanis2024}.

The $s+p$ state specific to $p$wM responds non-trivially to the applied field by turning into a non-unitary $s + p + ip'$ mixed symmetry state. 
These unconventional states are robust against magnetic impurities if the impurity magnetic moments can be oriented perpendicular to the spin polarization.
The superconductivity is not incompatible with coplanar magnetism \cite{Stadel2022,Maeda2025}. 
We expect the mixed parity unconventional character of the order parameter to have a direct impact on the properties of the Josephson junctions with NR-IS serving as a weak link \cite{Fukaya2025,Fukaya2025a}.

The NR-IS may possess a relativistic SOC.
In SOC-IS as long as Rashba and Dresselhaus terms are small, the formally infinite Pauli limiting field becomes finite but still anomalously large \cite{Wang2021,Haim2022,Harms2026}. Strict ``Isingness'' is not required for what is discussed in this Letter. 
All physics remains the same, only somewhat weakened. For our NR-IS the same is even much more true: potential Ising SOC will only enhance the effect, and Rashba or Dresselhaus terms will be much weaker than the nonrelativistic Ising splitting.

The work opens an exciting possibility for unconventional pairing such as coexisting $s$-wave singlets and $h$-wave triplets in an experimentally realized LaFeAs$_{0.6}$P$_{0.4}$ NR-IS.

\begin{acknowledgments}
We are indebted to M. Christensen for pointing our attention to the work \cite{Stadel2022}.
MK acknowledges the financial support from the Israel
Science Foundation, Grant No. 2665/20. LŠ acknowledges funding from the ERC Starting Grant No. 101165122 and Deutsche Forschungs-
gemeinschaft (DFG) grant no. TRR 288 – 7422213477 (Projects A09 and B05).
 IM was supported by the Office of Naval Research through grant \#N00014-23-1-2480.
\end{acknowledgments}


%

\end{document}